\documentclass[aps,prl,twocolumn,superscriptaddress,english]{revtex4-2}
\usepackage[utf8]{inputenc}
\usepackage{amsmath}
\usepackage{amssymb}
\usepackage{graphicx}
\usepackage{xcolor}
\usepackage{siunitx}
\usepackage{lipsum}
\usepackage{orcidlink}
\usepackage{mathrsfs}
\usepackage{lineno}
\usepackage{booktabs} 

\begin{document}

\title{High-pressure behavior of the low-frequency mode of water} 

\author{Maria Rescigno}
\email{rescigno@ill.fr}
\affiliation{Institut Laue-Langevin, 71 Avenue des Martyrs, 38042, Grenoble, France}
\author{Guillaume Radtke}
\affiliation{Institut de Min\'{e}ralogie, de Physique des Mat\'{e}riaux et de Cosmochimie (IMPMC), Sorbonne Universit\'{e}, CNRS UMR 7590,  MNHN, 4, place Jussieu, Paris, France}
\author{Umbertoluca Ranieri}
\affiliation{Centro de Física de Materiales (CFM-MPC), CSIC-UPV/EHU, Donostia/San Sebastián 20018, Spain}
\author{Richard Gaal}
\affiliation{Laboratory of Quantum Magnetism, Institute of Physics, \'{E}cole Polytechnique F\'{e}d\'{e}rale de Lausanne, CH-1015 Lausanne, Switzerland}
\author{Luigi Paolasini}
\affiliation{European Synchrotron Radiation Facility, 71 Avenue des Martyrs, 38000 Grenoble, France}
\author{Andrea Piovano}
\affiliation{Institut Laue-Langevin, 71 Avenue des Martyrs, 38042, Grenoble, France}
\author{Ferdinando Formisano}
\affiliation{CNR-IOM \& INSIDE@ILL, Operative Group in Grenoble, 71 avenue des Martyrs, F-38042 Grenoble, France}
\affiliation{Institut Laue-Langevin, 71 Avenue des Martyrs, 38042, Grenoble, France}
\author{Paola Giura}
\affiliation{Institut de Min\'{e}ralogie, de Physique des Mat\'{e}riaux et de Cosmochimie (IMPMC), Sorbonne Universit\'{e}, CNRS UMR 7590,  MNHN, 4, place Jussieu, Paris, France}
\author{Livia Eleonora Bove}
\affiliation{Institut de Min\'{e}ralogie, de Physique des Mat\'{e}riaux et de Cosmochimie (IMPMC), Sorbonne Universit\'{e}, CNRS UMR 7590,  MNHN, 4, place Jussieu, Paris, France}
\affiliation{Dipartimento di Fisica, Sapienza Universit\`a di Roma, Piazzale Aldo Moro 5, 00185 Roma, Italy}
\affiliation{Laboratory of Quantum Magnetism, Institute of Physics, \'{E}cole Polytechnique F\'{e}d\'{e}rale de Lausanne, CH-1015 Lausanne, Switzerland}

\begin{abstract}
The low-frequency (LF) mode observed in the THz spectrum of liquid water near 10~meV has been known for decades, yet its microscopic origin remains debated. Leading interpretations assign it to a transverse acoustic-like excitation, a transverse optic-like mode, or a mode-mode coupling mechanism. We address this question by probing the collective dynamics of H$_2$O and D$_2$O at room temperature and pressures from 0.1 to 1~GPa using high-resolution (1.6~meV) inelastic X-ray scattering (IXS). The LF mode remains heavily damped, close to critical damping, throughout the explored range, while its frequency hardens markedly with pressure, much more than expected for a transverse acoustic mode. Comparison with IXS measurements and \textit{ab initio} simulations of crystalline Ice VI reveals a cluster of weakly dispersive modes in the same energy range, suggesting that the low-frequency mode in liquid water reflects a mixed character arising from the superposition of several branches rather than a single well-defined excitation, providing a coherent microscopic picture of water's anomalous THz dynamics under compression.
\end{abstract}

\maketitle

Water's anomalous physical properties, essential to life, originate from the directional hydrogen-bond network imposed by the polar H$_2$O geometry \cite{debenedetti2003}. This network governs not only self-dynamics \cite{bove_translational_2013,ranieri_dynamical_2016,amann-winkel_x-ray_2016} and the exceptionally rich phase diagram \cite{Plastic_water,iceXXI}, but also the collective, phonon-like excitations, which differ markedly from those of simple liquids \cite{revirew_ruocco,Bove_Hg,Bove_K}. Pioneering theoretical \cite{rahman_simulations} and experimental \cite{bosi1978observation,teixeira1985observation} studies revealed a rich THz spectrum: at low wavevectors  ($Q \lesssim 4$~nm$^{-1}$), a single longitudinal acoustic mode propagates at the adiabatic sound velocity ($c_s \sim 1500$~m/s), while at larger wavevectors the apparent velocity rises to $\sim$3200~m/s, close to that of hexagonal ice~I\textit{h} \cite{iceIh}.
Initially termed ``fast sound'' by analogy with disparate-mass binary alloys \cite{Smorenburg1996,FastSound_LiTl_LiPb_1994}, this feature is now understood as a viscoelastic crossover \cite{normal_fast_sound,sciortino_sastry_MD,nardone_BLS,Masciovecchio2,masciovecchio_fast_sound,SANTUCCI2005941}: as the structural relaxation time $\tau$
enters the experimental window with increasing $Q$, the liquid crosses from viscous to solid-like behavior, with excitation velocities approaching those of ice~I\textit{h}.\\\\ 
A second, lower-energy feature ($\sim$10~meV) appears consistently above $Q \approx 4$~nm$^{-1}$. This low-frequency (LF) excitation has been reported by coherent inelastic neutron scattering \cite{bosi1978observation,teixeira1985observation,petrillo1,petrillo2,cunsolo_prb_2012}, high-resolution IXS \cite{revirew_ruocco,cunsolo,U_Balucani_1996,normal_fast_sound,coll6}, incoherent INS \cite{BELLISSENTFUNEL1991213}, and Raman spectroscopy \cite{raman_mode}, yet its microscopic origin has remained a longstanding open question \cite{cunsolo}. Three competing interpretations have been proposed: (i)~a transverse optical-like excitation \cite{normal_fast_sound,iceIh}, analogous to weakly dispersing branches in amorphous ices \cite{ixs_HDA,ixs_iceXII,ixs_vHDA,ics_HDA_LDA_Ic} and disparate-mass alloys \cite{Taras_Bryk_2002,Bove_2007,Bove_2011}; (ii)~a transverse acoustic-like mode \cite{coll2}, rendered accessible by partial longitudinal-transverse symmetry mixing \cite{coll6,coll7}, as observed in disordered tetrahedral networks and glasses \cite{GeO2_Bove_2005,Pilla_2004}; or (iii)~a mode-mode coupling feature arising from an avoided crossing between a linearly dispersing density fluctuation and a dispersionless mode \cite{petrillo1,petrillo2}.
Pressure is a powerful discriminator: by tuning the density and distorting the tetrahedral hydrogen-bond network, it directly links structural reorganization to collective dynamics. Yet high-resolution IXS studies of water under pressure remain scarce and, to our knowledge, have not exceeded a few kilobars \cite{coll2,monaco_IXS_HP_1999}, mainly owing to the small sample volume in diamond anvil cells. Intermediate-resolution studies addressed the viscoelastic crossover \cite{krisch_PRL_2002,santoro_PRL_2011,ranieri_dynamical_2016} but lacked the energy resolution needed to access the LF mode directly.\\\\
Here we present high-resolution (1.6~meV) IXS measurements performed on the ID28 beamline \cite{Bosak02072024} at ESRF (Grenoble, France) on H$_2$O and D$_2$O at room temperature and pressures from 0.1 to 1~GPa, approaching the ice~VI crystallization boundary (the experimental phase diagram and the thermodynamic path followed are reported in figure S5 in Supplementary Material \cite{suppMat}). By exploiting a specifically developed large-volume piston-cylinder cell that overcomes the sample-size limitation of earlier work, and by comparing the liquid excitations directly with the ice~VI phonon dispersion from \textit{ab initio} simulations, we show that the LF mode is inconsistent with assignment to a single, purely acoustic or optic branch.\\
\begin{figure}
\includegraphics[width=0.35\textwidth]{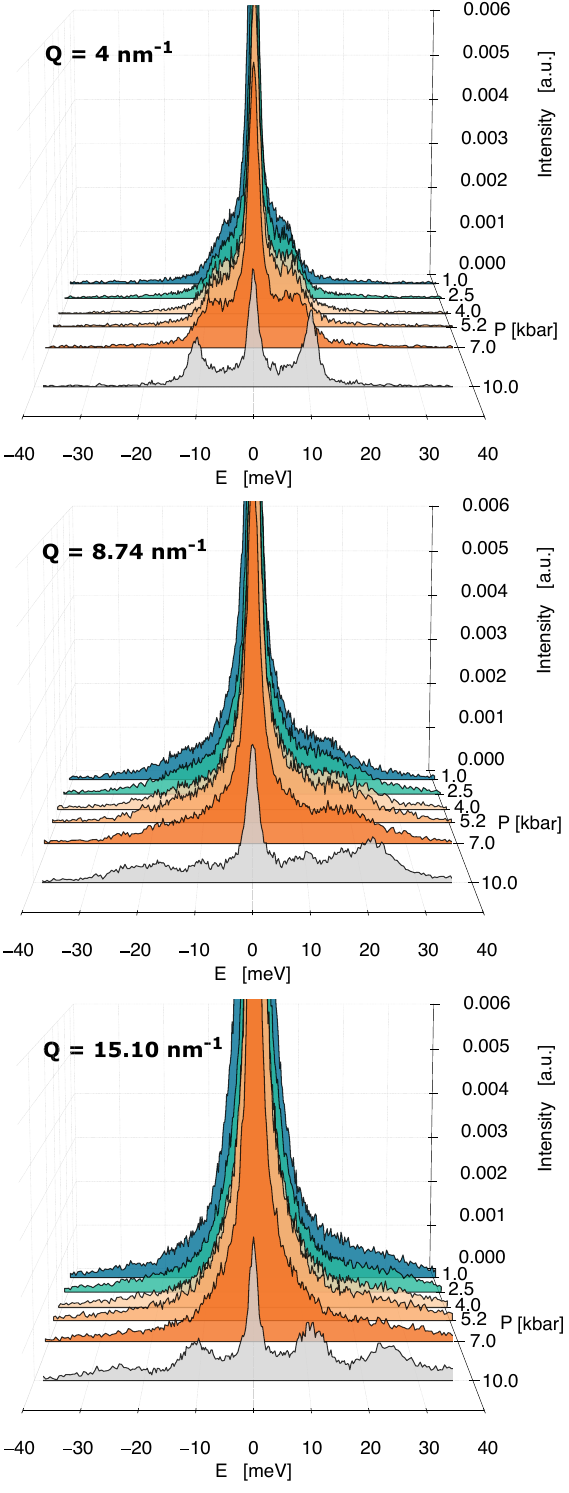}%
\caption{Inelastic X-Ray Scattering spectra of the D$_2$O sample at 300 K for selected Q values for all the probed pressures, from 1 to 7 kbar for the liquid sample and at 10 kbar on a solid (Ice VI) sample.\label{fig:spectra}}%
\end{figure}\\
\noindent
At all pressures in the liquid phase, the measured spectra (figure~\ref{fig:spectra}) are fitted using a Bayesian approach \cite{DeFrancesco2016} with a sum of a central Lorentzian peak and $k$ damped harmonic oscillator (DHO) functions convolved with the experimental resolution. With this method, the number of DHO components $k$ is
itself a parameter of the fit. The $DHO(Q,\hbar\omega)$ function --multpiled by the detailed balance-- is expressed as:
\begin{equation*}\begin{split}
DHO(Q,\hbar\omega) = \frac{A(Q)}{\pi}\frac{\beta}{1-e^{-\beta\hbar\omega}}\\
\frac{\Gamma(Q)\hbar\omega}{[(\hbar\omega)^2-(\hbar\Omega(Q))^2]^2+(\Gamma(Q))^2(\hbar\omega)^2}\label{eq:DHO}\end{split}\end{equation*}
where A(Q) is an intensity factor, $\Gamma$(Q) is the broadening of the modes and $\Omega$(Q) is the resonant frequency and $\beta$ is 1/K$_B$T. The choice of the DHO model to describe the excitations, instead of generalized liquid models \cite{PhysRevE_guarini_models} as the visco-elastic one often used to describe the inelastic spectrum of liquid water \cite{revirew_ruocco}, is dictated by the need to compare with the analysis of the crystalline sample. For each of the 56 spectra, including Q = 4 nm$^{-1}$, this analysis \cite{DeFrancesco2016} gives a posterior probability P($k$=2$\mid$y) $>$ 0.999, with $k$=1 or 2 allowed (more details on the methodology are reported in section S2-B of the Supplementary Material \cite{suppMat}). 
Examples of fit and best fit parameters are shown in figure \ref{fig:fitsD1}, while a comparison with a least-square fitting model with fixed number of DHO functions is shown in section S2-C, where it can be seen how---especially in the intermediate Q range---the fit quality improves when two DHO modes are considered. Additional statistical tests (section S2-C.2) also confirm this observation.\\\\
Two modes are observed: a propagative longitudinal acoustic (LA) mode (blue shaded area) and a heavily damped weakly dispersive low-frequency (LF) mode (green shaded area). The energies $\hbar\Omega$ of the two modes in the liquid phase, are shown in figure \ref{fig:band_struct}c for all pressures. The evolution of the two excitations with pressure offers a direct route to identifying their microscopic origin.\\\\
The LA mode behaves as expected for a propagating sound mode: its frequency follows the high-frequency sound velocity c$_{\infty}$, which increases from 3150 to 3540 m/s between 1 and 7 kbar (Table S2), and its damping grows approximately as Q$^2$ at low Q, close to the hydrodynamic behaviour, and more slowly at higher Q (Supplementary Section S2-B.1).
\begin{figure}[t]
\includegraphics[width=0.48\textwidth]{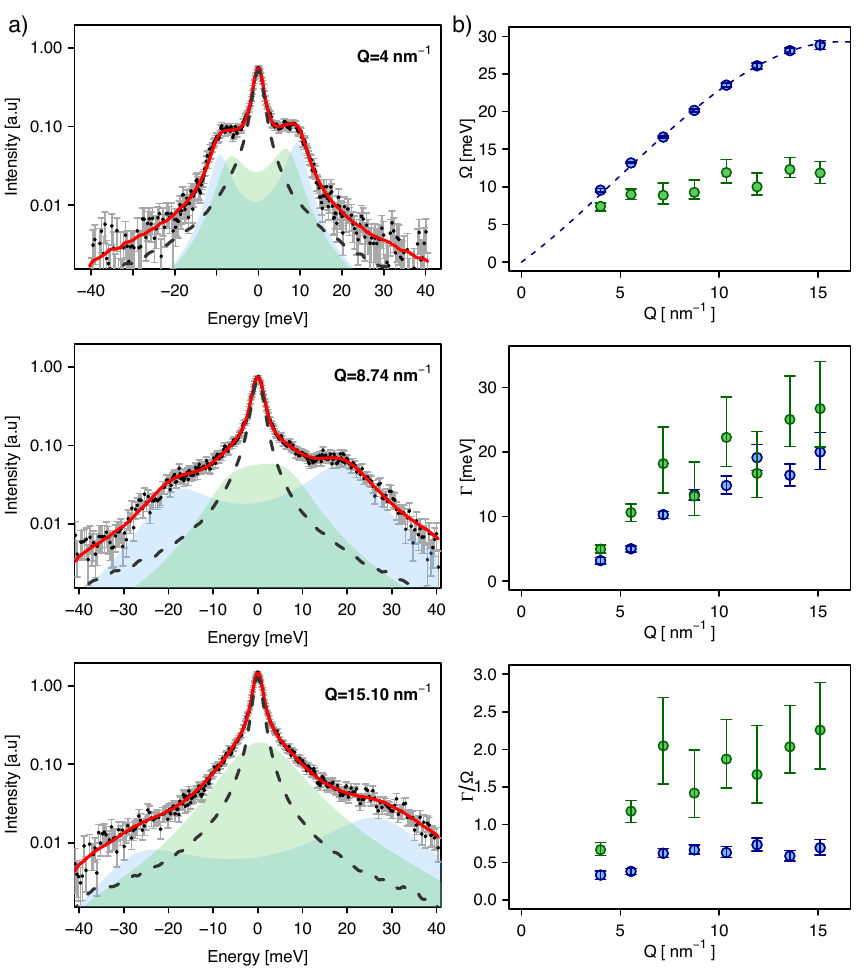}%
\caption{a) Examples of IXS spectra of D$_2$O at 300 K and 7 kbar, and their best Bayesian fits (red solid line) at three selected Q values (4, 8.74 and 15.10 nm$^{-1}$). The black dashed line is the elastic peak convolved with the experimental resolution, the green and blue shaded areas represent respectively the low and high frequency modes. b) Frequency $\Omega$, width $\Gamma$ and damping factor $\Gamma/\Omega$ for the LA (blue) and LF (green) modes.\label{fig:fitsD1}}%
\end{figure}
\begin{figure*}[t]
\includegraphics[width=1\textwidth]{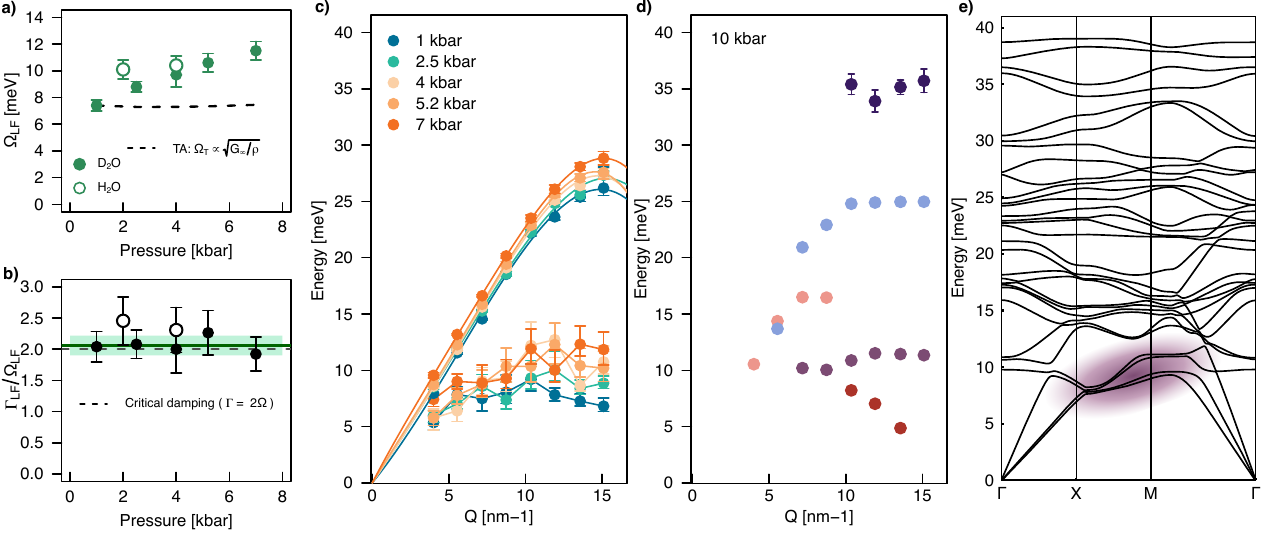}%
\caption{a) Comparison of the scaled calculated $\sqrt{G_{\infty}/\rho}$ and experimental $\Omega_{LF}$ as a function of pressure for D$_2$O (full circles) and H$_2$O (empty circles). Literature values used for the derivation are reported in supplementary table S5, the prediction is normalized to the value measured at 1 kbar. b) $\Gamma_{LF}/\Omega_{LF}$ as a function of pressure for D$_2$O compared with its mean value and the reference for critical damping. For both a) and b) the values are obtained as Q average with Q $\geq$ 10 nm$^{-1}$. c) Experimental dispersion of the LA and LF modes in liquid D$_2$O for pressures between 1 and 7 kbar. d) Experimental dispersion of powder Ice VI at 10 kbar obtained by fitting the IXS spectra with multiple DHO functions. Different colors refer to different modes observed in the IXS spectra. e) Low-energy phonon branches calculated using Density functional theory for the lowest-energy configuration obtained for Ice VI. The high symmetry points X and M are located approximately 5 and 10 nm$^{-1}$, respectively, from $\Gamma$ along the plotted path. The purple shaded area corresponds to the energy range where the low-frequency mode is observed in liquid water.
\label{fig:band_struct}}%
\end{figure*}\\\noindent
The LF mode behaves differently. Its damping ratio is pressure independent,
$\Gamma_{LF}/\Omega_{LF} = 2.05 \pm 0.12 $ (plateau value for Q $\geq$ 10 nm$^{-1}$, Fig. 3b), i.e. the mode remains close to critical damping over the whole range, while its frequency increases by 55 ± 12\% between 1 and 7 kbar (Fig. 3a). This provides a direct test of a transverse-acoustic assignment, since for a TA mode the frequency is set by the infinite-frequency shear modulus, $\Omega_T = \sqrt{G_{\infty}/\rho}\,$Q. In water at room temperature G$_{\infty}$ has been derived from combined viscosity and Brillouin measurements up to 1.5 GPa \cite{bulk_visc}: between 1 and 7 kbar it increases by less than 20\%, so that $\sqrt{G_{\infty}/\rho}$ is constant within about 1\% (dashed line in Fig. 3a). The data used for this derivation are reported in table S5 in section S2-D of the Supplementary Materials. The observed hardening of the LF mode is therefore incompatible with a purely transverse acoustic character.\\\\
\noindent
In addition to D$_2$O, a fully hydrogenated sample was also measured at 2 and 4 kbar to estimate the isotopic effect (see section S2-B.2 of Supplementary Material, where a direct comparison at 4 kbar is shown in figure S13). The weighted average over Q of the ratio of the frequencies $\Omega^D_{LF}/\Omega^H_{LF}$ of the mode is 0.92 $\pm$ 0.06, i.e. it scales as $\sqrt{m_{H_2O}/m_{D_2O}}$, suggesting that the mode is associated with the movement of the full molecules. This observation argues against a purely optic nature of the mode.\\\\
Inelastic Neutron Scattering (INS) experiments were also performed at 5 kbar on the IN8 triple axis spectrometer at the Institut Laue-Langevin (ILL) \cite{expIN8}, the data and details on the analysis are presented in the section S3 of Supplementary Material. Despite the limited intensity of the sample compared to the sample environment and the limited dynamical range of the experiment, the data suggest that the mode-mode interaction interpretation proposed in references \cite{petrillo1,petrillo2} does not satisfactorily describe the low Q (Q$<$4 nm$^{-1}$) behavior of liquid D$_2$O as no indication for a low Q flattening of the high frequency mode is observed. Recent sub-meV resolution IXS experiments on water under ambient conditions by Ishikawa et al. \cite{Ishikawa_quasi_inter_2021} show no evidence for a second mode for Q $<$ 4.2 nm$^{-1}$, thereby ruling out Sacchetti's mode-mode interaction model. Their measurements, extending over a broader energy range than the present INS data, independently confirm the same conclusion. The authors attribute the residual low-Q inelastic intensity to the coupling between the quasi-elastic and acoustic modes. Although the statistical quality and dynamical range of our INS data do not allow a quantitative test of the Ishikawa model, the residual intensity observed in our spectra is in qualitative agreement with this interpretation.\\\\
Finally, as the D$_2$O sample is compressed to 10 kbar, the sample crystallizes in a polycrystalline Ice VI phase \cite{iceVI}---the tetragonal hydrogen-disordered phase of Ice stable in the 10-20 kbar range at room temperature. As the system transitions to its crystalline phase, the IXS spectra are strongly modified and the appearance of multiple sharp bands is observed (gray spectra in figure \ref{fig:spectra}).
Density functional theory calculations have been performed to model these spectra following closely the procedure 
described in Ref.~\cite{Radtke_2024} (see section S1 in Supplementary Material). In particular, hydrogen disorder has been accounted for through a statistical average of 45 symmetrically distinct ice-rule-allowed configurations compatible with the tetragonal unit cell containing 10 water molecules~\cite{Knight_2005,Kuo_2006,Komatsu_2016}. 
Although the combination of a polycrystalline sample and the use of large momentum transfers leads to the simultaneous interception of many Brillouin zones and therefore hinders a straightforward assignment of the observed features, experimental spectra are qualitatively well reproduced (see supplementary figure~S20). Most importantly, both experimental and theoretical spectra show a weakly dispersing excitation at about 10 meV, which matches the low-frequency mode observed in the liquid phase. The experimental IXS spectra can be analyzed similarly to the liquid data (section S2-E in Supplementary Material). 
At Q = 4 nm$^{-1}$ only one peak is observed, the longitudinal acoustic mode, with an average sound velocity of (4027 $\pm$ 11) m/s, in good agreement with literature Brillouin spectroscopy values \cite{acoustic_velocity_iceVI}. At higher Q values different modes enter the probed dynamical range, with frequency $\Omega(Q)$ shown in figure \ref{fig:band_struct}d. As already qualitatively observed, a weakly dispersive mode is measured at about 10 meV. Such observed mode can be compared with other literature data on powder crystalline ice samples: hexagonal ice Ih \cite{iceIh}, ice XII \cite{ixs_iceXII}, stacked disordered ice Isd \cite{ics_HDA_LDA_Ic} and Ice IX \cite{ixs_HDA} where a similar mode at slightly lower frequencies is reported. Recent studies on polycrystalline Ice Ih \cite{DeFrancesco_2023} also observe a LF mode in a wide Q range, the authors interpret it as a collective mode undergoing an acoustic-to-optic crossover with increasing momentum transfer and suggest that the LF mode of liquid water could arise from a similar behaviour. Single crystal studies, which would allow to firmly assign the nature of the observed excitations, are limited to the case of hexagonal ice (figure 11 of reference \cite{single_crystal_Ih}), where both a TA and a TO mode are measured in the same energy region. Finally, figure~\ref{fig:band_struct}e shows the low-energy phonon branches calculated for the lowest-energy ordered configuration compatible with Ice VI, which displays the symmetry of the space group $P_c$ consistent with previous studies~\cite{Knight_2005,Kuo_2006,Komatsu_2016}. It appears that weakly dispersive low-energy optical modes, originating from the rigid out-of-phase vibration of the two isolated hydrogen-bonding sub-networks, occur around 10 meV and progressively mix with the acoustic branches in this energy range.\\\\
In conclusion, high-resolution IXS on H$_2$O and D$_2$O at 0.1--1~GPa provides a comprehensive picture of the collective dynamics of water under compression and sheds new light on the longstanding debate over the microscopic origin of the low-frequency mode. Two excitations are observed throughout the liquid phase for Q$\geq$4 nm$^{-1}$: a high-frequency LA mode with sound velocities consistent with prior literature \cite{krisch_PRL_2002,santoro_PRL_2011,ranieri_dynamical_2016}, and a weakly dispersive, heavily damped LF mode near 10~meV. The pressure degree of freedom, largely unexplored at this energy resolution, proves to be a decisive discriminator among the competing interpretations. Three independent and mutually consistent observations converge to rule out a single-branch assignment. First, the LF frequency increases by about 55\% between 1 and 7 kbar, whereas the shear modulus of water, which would set the frequency of a transverse acoustic mode, changes by less than 20\%.
Second, the isotopic frequency ratio $\Omega^D_{LF}/\Omega^H_{LF} \approx \sqrt{m_{H_2O}/m_{D_2O}}$ is consistent with a mode involving the motion of whole molecules rather than protons alone, ruling out a transverse proton-dominated optic-like assignment. Third, and most directly, the comparison with IXS measurements and \textit{ab initio} simulations of ice~VI reveals a cluster of weakly dispersive optical branches near 10~meV, originating from the out-of-phase vibration of the two hydrogen-bonding sub-networks, which progressively hybridize with the acoustic branches in this energy range. The liquid LF mode naturally maps onto this manifold of overlapping excitations rather than onto any single branch. Taken together, these results strongly support the conclusion that the LF mode of liquid water is not a well-defined elementary excitation of purely acoustic or optic character, but rather a superposition of transverse acoustic and transverse optic branches whose mixed nature reflects the topological complexity of the tetrahedral hydrogen-bond network. This complexity is robust: it survives the substantial structural reorganization induced by compression up to the crystallization boundary of ice~VI, underscoring that the anomalous THz dynamics of water are an intrinsic consequence of the versatility of the hydrogen bond, whose tetrahedral coordination persists under pressure through the interpenetration of open ring motifs rather than their disruption.

\section{acknowledgments}
We acknowledge the European Synchrotron Radiation Facility (ESRF) for provision of synchrotron radiation facilities under proposal number HC3412. We thank the Institut Laue-Langevin (ILL) for provision of beamtime on the IN8 triple axis spectrometer (experiment number DIR-317), and Claude Payre for assistance with the high pressure sample environment. We thank Frédéric Caupin for participating in the neutron scattering experiment. L.E.B., F.F. and M.R. acknowledge the financial support by the European Union - NextGenerationEU (PRIN N. 2022NRBLPT). L.E.B and M.R. acknowledge the ANR-23-CE30-0034 EXOTIC-ICE. L.E.B. and R.G. acknowledge the financial support by the Swiss National Fund (FNS) under Grant No. 212889. U.R. acknowledges support from the Diputación Foral de Gipuzkoa through Grant 2025-FELL-000013-01. This work was granted access to the HPC resources of IDRIS under the allocation AD010910820R3 made by Grand Equipement National de Calcul Intensif (GENCI).

\nocite{Giannozzi_2017,Schlipf_2015,Perdew_1996,kuhs_structure_1984,Baroni_2001,Sturm_1993,Baron_2014,Herrig_EOS_D2O,viscosity_D2O}
\bibliography{bibliography_new}

\end{document}